**Magnetoelectric coupling induced by interfacial orbital reconstruction**

*Bin Cui, Cheng Song\*, Haijun Mao, Huaqiang Wu, Fan Li, Jingjing Peng, Guangyue Wang, Fei Zeng, and Feng Pan*

B. Cui, Dr. C. Song, H. J. Mao, F. Li, J. J. Peng, G. Y. Wang, Dr. F. Zeng, and Prof. F. Pan

Key Laboratory of Advanced Materials (MOE), School of Materials Science and Engineering, Tsinghua University, Beijing 100084, China

Dr. H. Q. Wu

Institute of Microelectronics, Tsinghua University, Beijing 100084, China.

E-mail: songcheng@mail.tsinghua.edu.cn

Keywords: orbital reconstruction, magnetoelectric coupling, interface

With the fast development of multifunctional devices, magnetoelectric (ME) coupling — a bridge between magnetic and electric parameters plays a vital role in the integration and manipulation of microelectronics.[1–7] The artificial interface between ferromagnetic (FM) and ferroelectric (FE) oxides provides an ideal arena for electric-field control of magnetism with reduced energy consumption. Following the extensive studies, the proposed mechanisms responsible for the ME coupling could be summarized as follows: i) the electrostriction introduces strain variation in the FM layer, changing its lattice and concomitant magnetic properties;[8] ii) the delicate modulation of carrier density by polarization reversal in FE field-effect transistor motivates the FM/antiferromagnetic (AFM) phase transition;[9,10] iii) the use of multiferroic materials provides a route to the electrical control of spin arrangement by FM/AFM exchange coupling.[11–13] However the realization of electrostriction and exchange coupling needs high-quality thick FE layer (even FE substrates) and rare multiferroic materials, respectively, while the carrier modulation only plays a significant role in the phase transition of materials near the FM/AFM critical points. Thus an elegant approach for the realization of ME coupling through a more generalized mechanism is intensely pursued.[14–16]

The orbital occupancy determines the magnitude and anisotropy of the inter-atomic electron-transfer interaction and hence exerts a key influence on the electronic structure and magnetic ordering.[17] The interfacial covalent bond based on orbital reconstruction connects two distinct oxides at their interface, giving rise to novel electronic structure and performance.[18,19] In FE field-effect transistor, the interfacial orbital reconstruction could be effectively manipulated by the spatial movement of charged ion during polarization reversal, which is the essential behavior in all the FE materials without thickness limitation, thus providing a more generalized way for the ME coupling through orbital reconstruction. Here we realize the ME coupling by orbital reconstruction under electric field in model FE/FM heterostructures $BaTiO_3/La_{2/3}Sr_{1/3}MnO_3$ (BTO/LSMO), which is expected to complete the mechanism accounting for ME coupling and to advance the development of orbital degree of freedom in the terrain of microelectronics.

High-quality BTO (10 u.c.)/LSMO ($t$ = 7, 10, 15, 25, and 50 u.c.) heterostructures were grown on $SrTiO_3$ (STO) substrates (Supporting Information Figure S1). The samples were then made into two types of devices (Supporting Information Figure S2 and Experimental section): a transistor device was used to carry out the transport measurements; while a film device was required for the synchrotron radiation measurements. A typical out-of-plane piezoresponse force microscopy (PFM) image of BTO (10 u.c.)/LSMO (15 u.c.) written by ±4 V is presented in **Figure 1**a, where the bright and dark regions stand for polarization upward and downward ($P_{up}$ and $P_{down}$), respectively. The in-plane compressive strain (Supporting Information Figure S3) guarantees the tetragonal BTO phase with good ferroelectricity. The as-grown BTO (outmost area of Figure 1a) exhibits a single domain with $P_{up}$ due to the polar discontinuity at the interface (Ref. 19). This is also concluded from the shift of local piezoelectric hysteresis and butterfly loops (Supporting Information Figure S4 and S5).

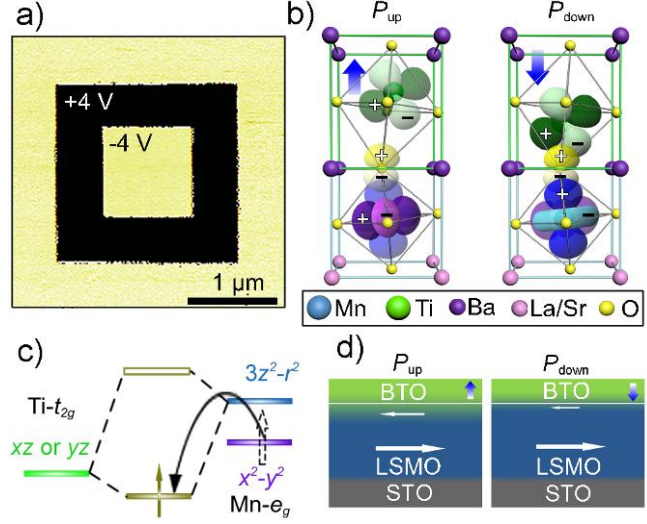

**Figure 1** a) Typical out-of-plane PFM images of BTO/LSMO heterostructure on $TiO_2$-terminated (001) STO substrate. b) Sketch for the interfacial orbital reconstruction under different polarization states. The favored orbitals are dense while the disfavored ones are washy. c) Schematic for the orbital hybridization between Ti, O, and Mn. (d) Schematic of the LSMO channel composited by interfacial and bulk layer under $P_{up}$ and $P_{down}$.

The FE displacement of BTO measured in butterfly loops is around 15 pm (close to the theoretical value of 12 pm),[20] which would induce the orbital reconstruction of Mn in the heterostructure as described in Figure 1b. A highly overlapped out-of-plane orbital component Mn-$d_{3z^2-r^2}$ and Ti-$d_{xz}$ (-$d_{yz}$) in the lattice under a $GdFeO_3$-type distortion would hybridize into a bonding orbital with lower energy level than that of the originally occupied Mn-$d_{x^2-y^2}$ (Figure 1c), strongly weakening the in-plane orbital occupancy.[21] The BTO in $P_{up}$ and $P_{down}$ states suppress and enhance the interfacial orbital hybridization, accompanied by the intensified and weakened Mn-$d_{x^2-y^2}$ orbital occupancy, respectively. The interfacial termination should not change the effect of polarization on interfacial orbital reconstruction as both the (La,Sr)O and BaO could supply O ions for the Ti-O-Mn bonding. Here the LSMO layer could be treated as two parallel channels of bulk and interface, as shown in Figure 1d. By altering the orbital occupancy at the interface, FE polarization could modulate the relative weight of the interfacial channel with preferential Mn-$d_{x^2-y^2}$ occupancy.[22,23] The interfacial layer behaves a spin arrangement antiparallel to the bulk one,[22] and its modulation under electric field is bound to influence the magnetic and transport properties of LSMO.

The temperature dependent resistance (*R-T*) curves of BTO/LSMO are displayed in **Figure 2**a. The Curie temperature ($T_C$) obtained from the metal-insulator transition temperature in Figure 2a and $\Delta T_C = T_C(P_{down}) - T_C(P_{up})$ are summarized as a function of LSMO thickness (*t*) in the left and right axis of Figure 2b, respectively. We firstly focus on the case of 7 u.c. LSMO. Compared with the case of $P_{up}$, the resistance of LSMO with $P_{down}$ BTO decreases and the metallic conductive region is distinctly extended, followed by a $T_C$ enhancement of ~17 K. The improvement of electric and magnetic performance under $P_{down}$ state originates from the thinner interfacial channel with the depopulated Mn-$d_{x^2-y^2}$ orbital ordering as shown in Figure 1d. The thickness of interfacial layer is estimated to be around 3 u.c. for both BTO/LSMO and LSMO/STO interfaces according to the direct observations and transport measurements.[23,24] Thus the metallic-insulating phase separation in LSMO of 7 u.c. is remarkable, accompanied by distinctly semiconducting conductivity at low temperature. The resistance difference between $P_{up}$ and $P_{down}$ states is gradually eliminated by increased *t* and the $\Delta T_C$ drops rapidly with $\Delta T_C \propto 1/t$, reflecting the field effect is limited to the interface. It should be mentioned here that the existence of dead layer at the interface between LSMO and substrates shrinks the effective bulk thickness and enlarges the amplitude of resistance and $T_C$ variations, especially in the ultrathin films (7–15 u.c.). Nevertheless it keeps unchanged during the polarization reversal, thus it is not considered in our work.

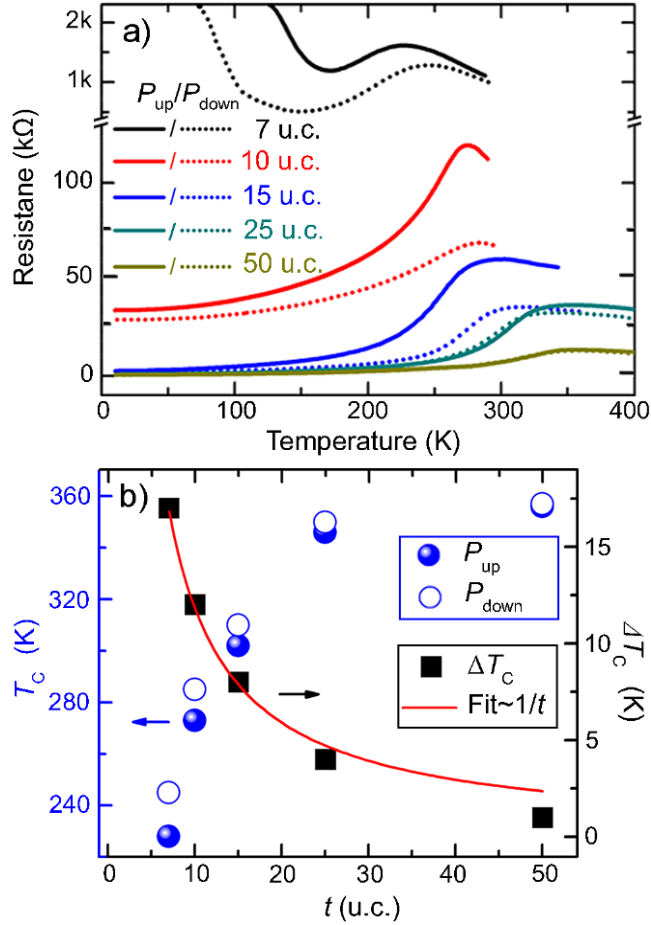

**Figure 2** a) *R-T* curves for BTO (10 u.c.)/LSMO (*t* = 7 u.c., 10 u.c., 15 u.c., 25 u.c., and 50 u.c.) transistor devices. The solid curves are for $P_{up}$ and dot ones for $P_{down}$. b) The $T_C$ of devices in $P_{up}$ and $P_{down}$ states (left axis) and their difference ($\Delta T_C$, right axis) as a function of LSMO thickness (*t*).

We now turn to directly explore the effect of the FE polarization on the interfacial magnetic properties by measuring the angular magnetoresistance (MR) of heterostructures at 10 K with magnetic field (*H*) rotating out-of-plane in **Figure 3**. The angle between magnetic field and normal direction of substrate is defined as $\theta$, where $\theta = 0°$ (180°) and 90° (270°) correspond to the field perpendicular to and in the film plane, respectively (Supporting Information Figure S2). Remarkably, besides the typical anisotropic MR for FM manganites, a novel MR with maxima for the in-plane magnetic fields (referred to as pMR) is superimposed in the normalized angular MR [Figure 3a (*H* = 5 T), 3b (*H* = 9 T) and Supporting Information Figure S6]. The most eminent feature observed here is that the heterostructures with up-polarized BTO exhibit stronger pMR peaks [pMR = ($R_{90°}$ − $R_{min}$)/ $R_{0°}$ ] than those with the down-polarized ones: the pMR values of $P_{up}$/$P_{down}$ are 0.070%/0.046% and 0.058%/0.034% for LSMO of 10 u.c. and 15 u.c. at 9 T, respectively. The pMR is highly related to the preferential $x^2$–$y^2$ occupation at the interfacial layer under tensile strain, where the transport properties are dominated by the bands around the Fermi energy ($3z^2$–$r^2$ orbital). The onsite energy on the $3z^2$–$r^2$ orbital increases with $\theta$ from 0°(out-of-plane) to 90°(in-plane), which would narrow the band around the Γ point and lead to a decreased velocity (increased resistance) for $\theta$ = 90°(pMR) under a magnetic field.[22] Then the variation of pMR peak with the switching of the FE polarization could be understood in terms of the orbital reconstruction: compared with the case of down-polarized state, the enhancement of Mn-$d_{x^2-y^2}$ orbital occupancy under up-polarized state favors the pMR.

The dependences of pMR on magnetic field and LSMO thickness are summarized in Figure 3c and 3d, respectively, where both the increased $H$ and decreased $t$ enlarge the pMR values. The height of the pMR peak rises with the enlarged magnetic field due to the band narrowing effect as shown in Figure 3c. The pMR value of $P_{up}$ state at 9 T rapidly reduces from 0.070% to 0.002% with the increase of LSMO thickness from 10 u.c. to 50 u.c. in Figure 3d, originating from the decreased relative weight of the interfacial layer. As a result, the difference between pMR values of $P_{up}$ and $P_{down}$ states is negligible in the thick film of 50 u.c., where the contribution of varied interfacial layer to the transport property is covered by the bulk signal. This electrical manipulation of transport properties is reversible (Supporting Information Figure S7).

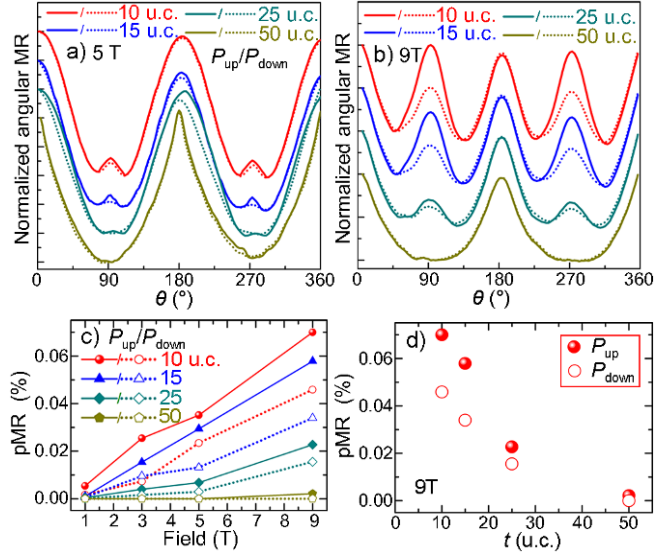

**Figure 3** a) and b), Normalized angular MR for BTO (10 u.c.)/LSMO ($t$ = 7 u.c., 10 u.c., 15 u.c., 25 u.c., and 50 u.c.) heterostructures at a magnetic field of 5 T and 9 T, respectively. The resistances are normalized with respect to $(R - R_{min})/(R_{max} - R_{min})$. c) The pMR values as a function of magnetic field for LSMO of different thicknesses. d) The dependence of pMR values under $P_{up}$ and $P_{down}$ on LSMO thickness at 9 T.

The electronic structures with different polarization states are then demonstrated by x-ray absorption spectroscopy (XAS) and x-ray linear dichroism (XLD) measurements at 300 K. **Figure 4**a is the XAS $L$-edge of Ti. The energy differences between the two main peaks of the $L_3$ ($\Delta E\ L_3$) and $L_2$ ($\Delta E\ L_2$) are displayed as a function of the LSMO thickness in Figure 4b. All the $\Delta E\ L_3$ and $\Delta E\ L_2$ values of samples with different polarization states and LSMO thicknesses are smaller than that of the reference sample of STO substrate, implying the valence lower than $Ti^{4+}$ (Ref. 21). The extra electrons in Ti ions arise from $Mn^{3+}$ through the orbital hybridization. Remarkably, the Ti in the $P_{down}$ state exhibits a lower valence than that in $P_{up}$ one with the same LSMO thickness, indicating a robust orbital hybridization for the $P_{down}$ case due to larger orbital overlap. On the contrary, the $L_3$ peak of Mn (marked by the black triangle) shifts toward higher energies as the polarization state changes from up to down (Figure 4c and 4d). Such a shift suggests a higher Mn valence in LSMO, exactly in consistent with the lower Ti valence.

The Mn-XLD curves in Figure 4e directly illustrate the variation of Mn orbital occupancy under different polarization states (more details in Supporting Information Figure S8). The area under XLD signal ($I_{ab}$–$I_c$) around $L_2$ peak (647.5 eV–660.0 eV) ($A_{XLD}$) highlighted in Figure 4e is used to represent the relative orbital occupancy: the more negative (positive) $A_{XLD}$ is, the larger relative occupancy of $d_{x^2-y^2}$ ($d_{3z^2-r^2}$) is.[25] The XLD results for LSMO thicker than 15 u.c. whose $T_C$ > 300 K are influenced by the FM signal and not shown here.[26] As LSMO grown on STO substrate is under tensile strain, the preferred orbital is $d_{x^2-y^2}$, demonstrated by the negative $A_{XLD}$ in the case of $P_{up}$ with the weak orbital hybridization. Conversely, the $A_{XLD}$ is positive and $d_{3z^2-r^2}$ is stabilized as the hybridization of out-of-plane orbital is enhanced with closer distance between Mn and Ti under $P_{down}$ state. The orbital occupancy altered by FE polarization are in line with the corresponding variation of pMR value based on Mn-$d_{x^2-y^2}$ occupancy, strongly supporting our understanding of ME coupling through the orbital reconstruction.

With the increase of LSMO thickness, the $\Delta E$ $L_3$ and $\Delta E$ $L_2$ of Ti XAS reduces while Mn $L_3$ peak position shifts to the higher energy direction, implying the gradual fall of Ti valence (Figure 4b) and the rise of Mn valence (Figure 4d), respectively, for both scenarios of $P_{up}$ and $P_{down}$. This can be explained by the fact that the in-plane preferential orbital occupancy introduced by STO substrate is reduced as the thickness of LSMO increases (Figure 4f). The dependence of valence and orbital occupancy on LSMO thickness in turn affirms that the modulation of electronic structure comes from the interface area. In previous studies, the role of 3d orbital in electric control of magnetism was investigated theoretically,[27–30] whereas our finding experimentally demonstrates the ME coupling induced by orbital reconstruction.

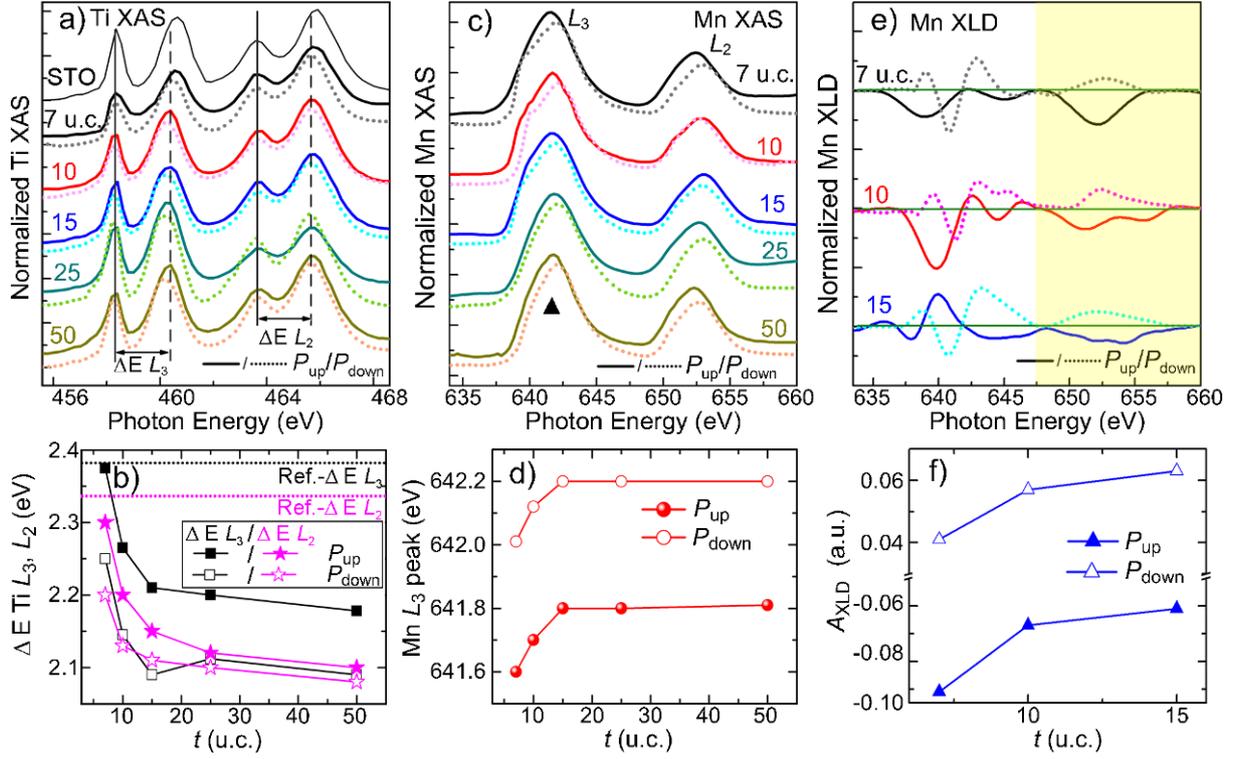

**Figure 4** Normalized a) Ti-XAS, c) Mn-XAS, and e) Mn-XLD for heterostructures with different LSMO thicknesses and polarization states (solid curves for $P_{up}$ and dot curves for $P_{down}$). The peak positions of Ti $L_3$ and $L_2$ edge are marked by solid and dashed straight lines and Mn $L_3$ peak is denoted by black triangle. b) The ΔE $L_3$ and $L_2$ in Ti-XAS, d) $L_3$ peak position in Mn-XAS and f) the $A_{XLD}$ of Mn-XLD as a function of LSMO thickness.

The role of exchange coupling in ME coupling could be easily excluded here by the absence of multiferroic materials. And the role of strain in ME coupling could be ruled out by the following aspects: i) the ultrathin FE layer weakens the clamping effect of strain produced by polarization; ii) the in-plane lattices of (001) BTO are almost the same in $P_{up}$ and $P_{down}$ states.[31] However, it is generally accepted that the FE polarization would tune the carrier density of LSMO channel, which might also contribute to the changes in magnetic and electric performance.[3,9,10] Meanwhile, the entanglement between orbital reconstruction and carrier modulation makes it difficult to absolutely separate the role of orbital in this ME coupling.

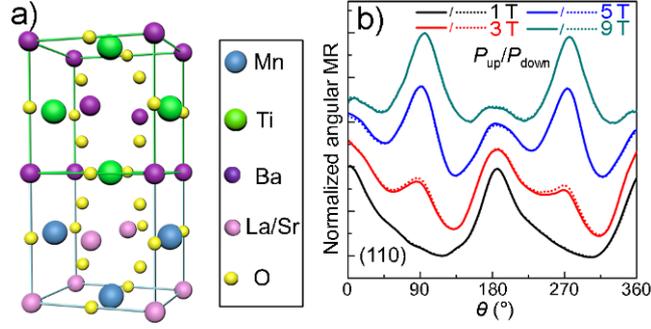

**Figure 5** a) Sketch for the atomic arrangement of (110) heterostructure. b) Normalized angular MR for (110) BTO/LSMO with different polarization states under magnetic fields ranging from 1 T to 9 T.

We then design a control experiment using heterostructures on (110) STO substrate, which was simultaneously prepared with the (001) BTO (10 u.c.)/LSMO (15 u.c.), to exclude the interference from the modulation of carrier density under polarization reversal to some extent. On the basis of the atomic arrangement along [110] direction in **Figure 5**a, the distance between Mn and Ti along the normal of substrate is enlarged without a connection between O. Thus the pMR peak of (110) heterostructure is higher than that of the (001) one due to the robust in-plane Mn orbital occupancy without the out-of-plane orbital hybridization, affirming the origin of pMR in turn. However, the FE poling of BTO (Supporting Information Figure S9) has a negligible influence on the pMR peak as shown in Figure 5b, because the orbital occupancy is stable in different polarization states. In addition, the small change of carrier density in $La_{2/3}Sr_{1/3}MnO_3$ (a robust and optimal FM phase) induced by polarization cannot obviously affect its electric properties without the assistance of orbital reconstruction (Supporting Information Figure S10). These observations in (110) heterostructure exclude the influence of carrier density variation on magnetotransport and reaffirm that the ME coupling observed in the (001) case is mainly caused by the orbital reconstruction rather than the carrier density change. Besides, we emphasize that the orbital reconstruction plays a dominant role in the ME coupling for the following reasons: i) the modulation of pMR signal is the direct reflection of orbital reconstruction under electric field;[22] ii) the delicate change of carrier density in the robust FM La$_{2/3}$Sr$_{1/3}$MnO$_3$ phase by polarization switching cannot obviously affect its magnetic and electric properties alone.[9,10,32,33]

In this work, two additional elements might enhance the ME coupling effect: phase separation and oxygen vacancy migration. The metallic-insulating phase separation is widely accepted in ultrathin LSMO. As the LSMO were grown on STO with subtle tensile strain, the separated insulating phase should be $x^2–y^2$ orbital ordering, which would increase the amount of controllable orbital and promote the manipulation effect additionally. Thus, it is reasonable to expect a larger ME coupling in Pr(Ca,Sr)MnO$_3$, and (La,Ca)MnO$_3$ with a stronger phase separation tendency.[16,33] On the other hand, the inevitable oxygen vacancies with positive charges in oxides would migrate towards and away from the interface in the $P_{up}$ and $P_{down}$ states, respectively.[34,35] As the Ti-O-Mn bonding is based on the orbital hybridization among Ti, O, and Mn ions,[36] the $P_{down}$ state with less oxygen vacancies at the interface would stabilize the bonding and orbital reconstruction, while the $P_{up}$ state shows the opposite behavior.

In summary, the interfacial orbital reconstruction of BTO/LSMO is driven by the shuttle displacement of Ti ion under FE polarization, displaying a direct and crucial role in manipulating the magnetism of heterostructures. With BTO polarized upward (downward), the orbital hybridization between Ti and Mn is suppressed (enhanced), followed by the interfacial preferential Mn-$d_{x^2-y^2}$ (Mn-$d_{3z^2-r^2}$) occupancy. The enhanced Mn-$d_{x^2-y^2}$ occupancy, rather than the changes in carrier density, in $P_{up}$ state increases the pMR values with suppressed magnetic and electric performances of FM metallic LSMO, while the suppressed Mn-$d_{x^2-y^2}$ occupancy of $P_{down}$ state does the opposite. This FE controlled bonding formation/distortion also provides a possible route to the reversible manipulation of novel magnetic state produced by the interfacial bonding.[21,23] Our findings not only present a broad opportunity to fill the missing member—orbital in the mechanism of magnetoelectric coupling, but also make the orbital degree of freedom straight forward to the application in microelectronic device.

**Experimental Section**

High-quality BTO (10 u.c.)/LSMO ($t$ = 7, 10, 15, 25, and 50 u.c.) heterostructures were grown in layer-by-layer mode on $TiO_2$-terminated STO substrates by pulsed layer deposition (PLD) from stoichiometric $La_{2/3}Sr_{1/3}MnO_3$ and $BaTiO_3$ targets. The LSMO was grown at 700 ℃ with an oxygen pressure of 200 mTorr while BTO was grown at 750 ℃ with an oxygen pressure of 4 mTorr. The growth was monitored *in situ* by RHEED (reflection high-energy electron diffraction) analysis, allowing the precise control of the thickness at the atomic level. Two types of devices were used in this work (Supporting Information Figure S2): a transistor device with a gate located in the vicinity of the channel was used to carry out the transport measurements in the physical property measurement system; while a film device of 2.5 mm × 5 mm was required for the XAS and XLD measurements in total electron yield mode at Beamline BL08U1A at Shanghai Synchrotron Radiation Facility.[33] A gate voltage of ±4 V was applied on the sample to switch the polarization and then removed before the measurements. The PFM measurements and ferroelectric switching were carried out by Cypher (Asylum Research).

**Acknowledgements**

We acknowledge Beamline BL08U1A in Shanghai Synchrotron Radiation Facility (SSRF) for XAS/XLD measurements. This work was supported by the National Natural Science Foundation of China (Grant Nos. 51322101, 51202125 and 51231004) and National Hi-tech (R&D) project of China (Grant no. 2014AA032904 and 2014AA032901).

## Supporting information

### Film growth

Epitaxial heterostructures BaTiO$_3$ (BTO)/La$_{2/3}$Sr$_{1/3}$MnO$_3$ (LSMO) with different BTO and LSMO thicknesses were grown on (001) and (110) SrTiO$_3$ (STO) substrates. The growth dynamics was investigated by monitoring the intensity variations of the RHEED (reflection high-energy electron diffraction) patterns on (001) STO substrate. Figure S1 shows the typically clear RHEED oscillations recorded during the growth of BTO (10 u.c.)/LSMO (15 u.c.) and BTO (10 u.c.)/LSMO (50 u.c.), suggesting a continued layer-by-layer growth of heterostructures and -(La,Sr)O-MnO$_2$-BaO-TiO$_2$- top interface termination. The growth rates of LSMO and BTO are 0.77 nm/min and 0.60 nm/min, respectively.

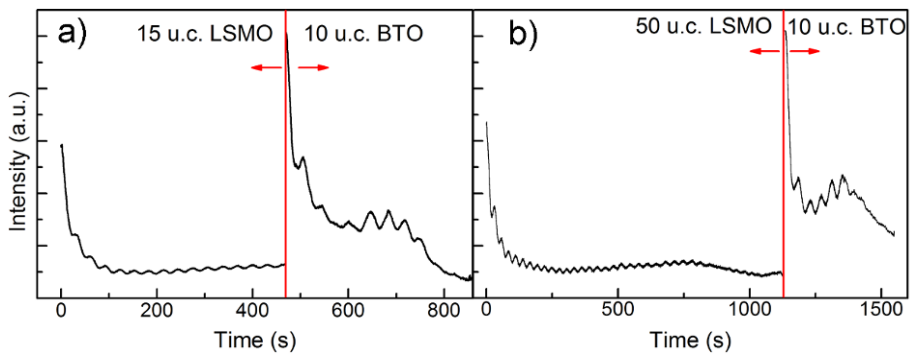

**Figure S1** RHEED intensity oscillations of a) BTO (10 u.c.)/LSMO (15 u.c.) and b) BTO (10 u.c.)/LSMO (50 u.c.) grown on (001) TiO$_2$-terminated STO substrate.

### Device preparation

There are two series of devices in this work. The transistor devices are used for the measurements of resistance versus temperature (*R-T*) curves and angular magnetoresistance (MR). The film was patterned into a transistor structure with a gate electrode located in the vicinity of channel by photo-lithography and wet etching as shown in Figure S2a. The effective size of the channel is 100 μm wide. A hard-baked photoresist covers the whole area except for the parts of electrodes and channel for electrical isolation between the gate and channel. Here the exposed channel is 70 μm wide, which is smaller than the whole one for preventing the leakage current at the side edge. A drop of silver paste is then used to connect the exposed channel and gate electrode. The rotation axis in the angular MR measurement is marked by the arrow in Figure S2a, while the magnetic field ($H$) is fixed. The film device with a large area of 2.5 mm × 5 mm is used for the measurements of x-ray absorption spectroscopy (XAS) and x-ray linear dichroism (XLD), where a metal-probe setup with a 50 μm tip was used to switch the polarization of the whole sample as shown in Figure S2b.[S1] A specific voltage is applied between gate electrode and heterostructure.

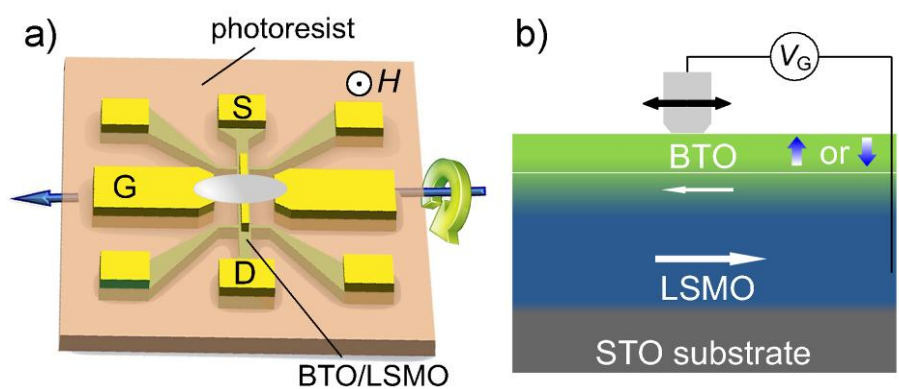

**Figure S2** a) The schematic of transistor device, where the source, drain, and gate electrodes are denoted as S, D, and G, respectively. The straight and bent arrows indicate the axis and direction of rotation, respectively. b) Sketch of the film device.

### Ferroelectricity of BaTiO$_3$ and its origin

The x-ray diffraction (XRD) taken from BTO (200 u.c.)/LSMO (50 u.c.) on TiO$_2$-terminated (001) STO substrate in Figure S3 reveals a $c$-axis orientation with no detectable impurity phase. The peak position of LSMO (002) at 47.92 ° demonstrates that a small in-plane tensile strain of ~0.52% exists in the LSMO layer. Meanwhile the 2-theta angle for BTO (002) peak is around 44.17 °, corresponding to $c$-axis lattice constant of 4.098 Å. This suggests that the BTO film undergoes in-plane compressive strain, producing a tetragonal ferroelectric (FE) phase.[S2]

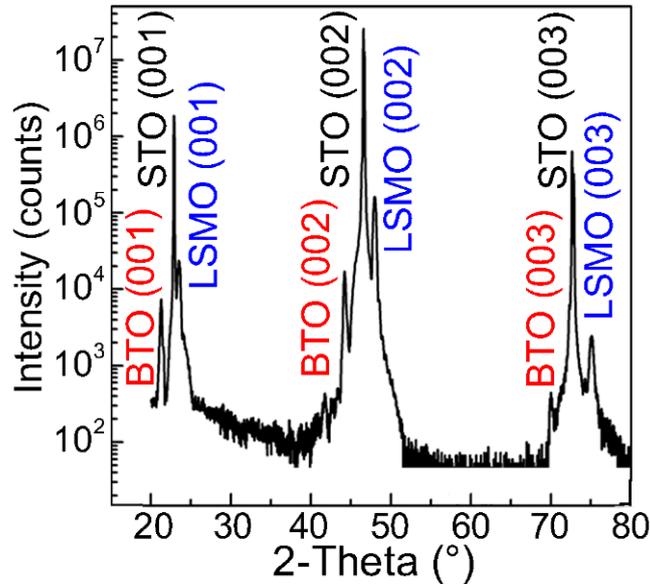

**Figure S3** XRD data of the BTO (200 u.c.)/LSMO (50 u.c.) heterostructure on (001) STO substrate.

The FE character of 10 u.c. BTO on 15 u.c. LSMO is confirmed by piezoresponse force microscopy (PFM) probe. The typical local piezoresponse hysteresis loops (Figure S4a) and well-shaped displacement-voltage (*D*–*V*) butterfly loops (Figure S4b) were measured in the as-grown area of out-of-plane PFM images in Figure 1a. The square hysteresis loops with sharply 180 ° FE switching in Figure S4a indicate that the heterostructure is of very high quality and possesses good FE properties. A positive voltage triggers the FE polarization towards the LSMO bottom layer, whereas a negative voltage polarizes the BTO upward. The FE displacement could be roughly estimated by *D*–*V* loops as shown Figure S4b, which exhibits a FE displacement of ~15 pm at ±4 V (close to the theoretical value of 12 pm[S3]). Both the piezoresponse hysteresis and *D*–*V* butterfly loops shift toward the negative direction, reflecting a spontaneously upward polarization, due to the particular polar discontinuity in -(La,Sr)O-$MnO_2$-BaO-$TiO_2$- at the top interface of the heterostructures.[S4] These results also suggest that a gate voltage of ±4 V is large enough to switch the polarization.

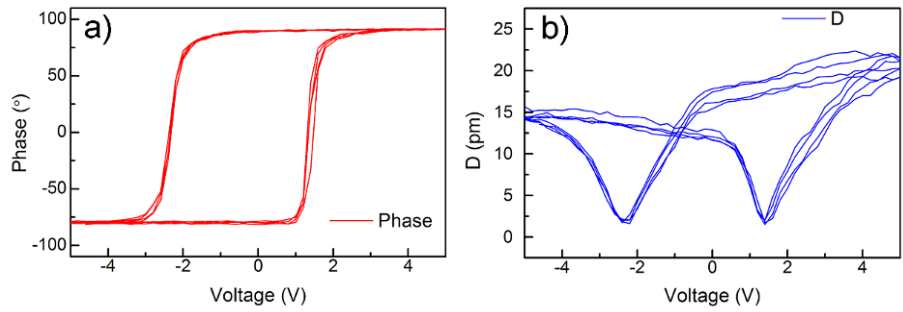

**Figure S4** a) Typical local piezoresponse hysteresis loops and b) displacement-voltage (*D–V*) butterfly loops of BTO (10 u.c.)/LSMO (15 u.c.) on (001) STO.

The polarization states of BTO (10 u.c.)/LSMO (15 u.c.) before and after poling with +4 V using a film device are studied by the out-of-plane PFM shown in Figure S5a and S5b, respectively. The contrast of the image is correlated with the orientation of the FE polarization, with the dark (bright) region indicating downward (upward) FE polarization. Strikingly, the outmost area of image is polarized downward in Figure S5b, reflecting the remnant polarization state of the film after +4 V poling, which is opposite to the case of as-grown one in Figure S5a. It is noteworthy that the phase contrast remains after repetitive measurements for many times. This result suggests that the approach we used to polarize the whole film is effective due to the high-quality of heterostructure under whole layer-by-layer growth mode, guaranteeing the *ex situ* XAS and XLD measurements.

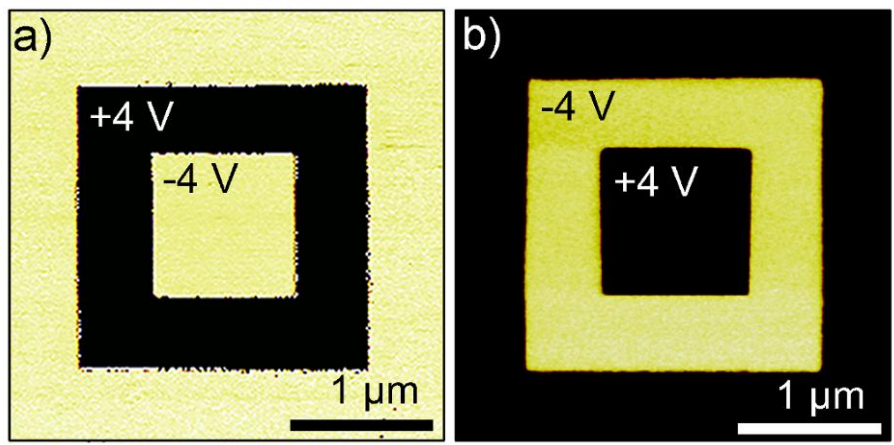

**Figure S5** Typical out-of-plane PFM image of BTO (10 u.c.)/LSMO (15 u.c.) a) before and b) after polarization with +4 V.

**The angular magnetoresistance for BTO/LSMO**

The raw angular MR measurement results by rotating the magnetic field out-of-plane are shown in Figure S6. The normalized curves in Figure 3a and 3b are obtained by setting the difference between the maximum and minimum points in Figure S6 to be one. In contrast to the classic anisotropic MR peaks at $\theta = 0°$ and $180°$ for ferromagnetic manganites, the MR peaks due to the interfacial $x^2 - y^2$ preferential orbital occupancy called pMR (in-plane angular magentoresistance) are located at the angles of $90°$ and $270°$ (Ref. S5). Both the decrease of LSMO thickness and increase of measured field would enhance the pMR peak, as the former increases the interfacial proportion and the latter induces the narrowing of band structure. More importantly, the pMR peak could be manipulated by the polarization of BTO, that the BTO of $P_{up}$ state exhibits larger pMR values than $P_{down}$ due to the enhancement of interfacial $x^2 - y^2$ occupancy.

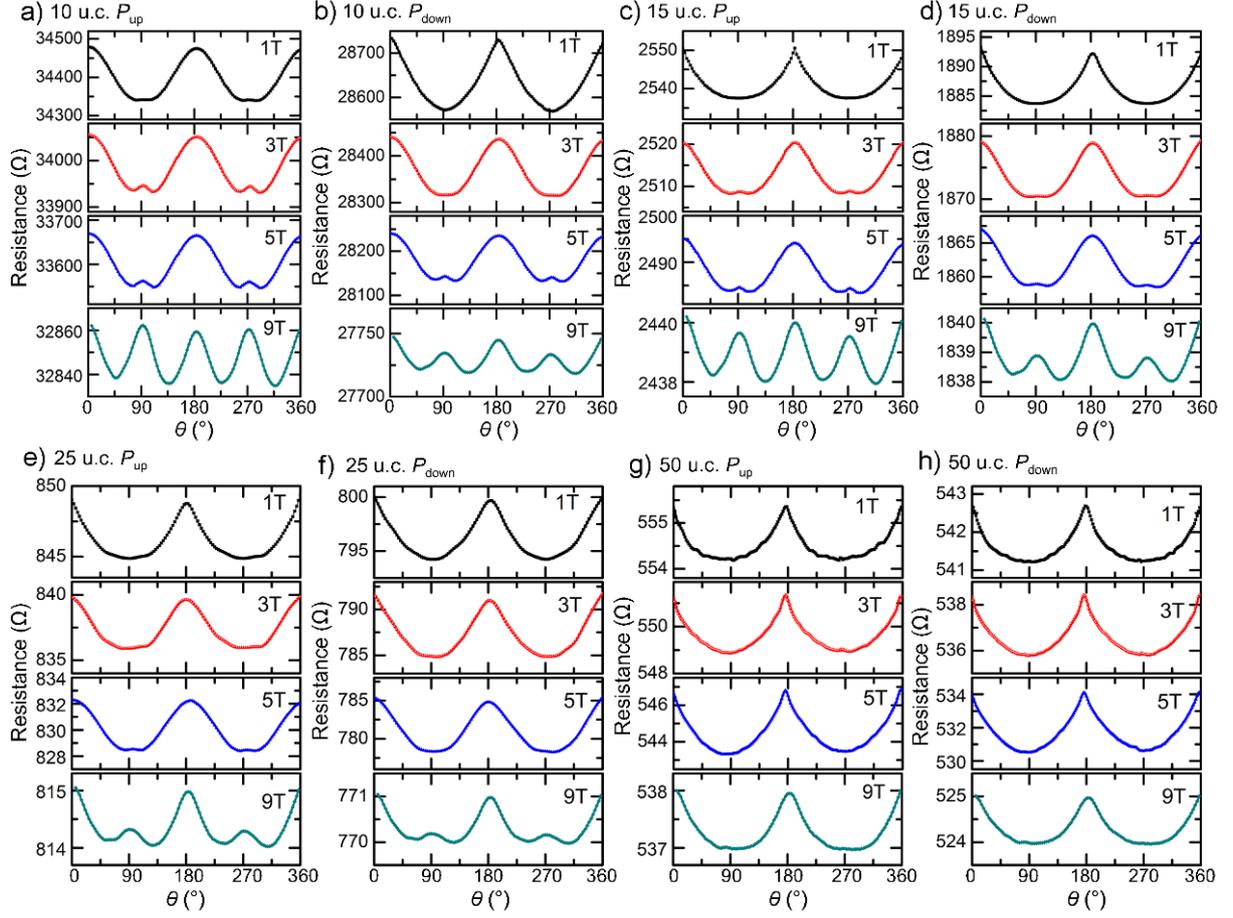

**Figure S6** The angular MR for BTO/LSMO with different LSMO thicknesses and BTO polarization states: 10 u.c. a) $P_{up}$ and b) $P_{down}$, 15 u.c. c) $P_{up}$ and d) $P_{down}$, 25 u.c. e) $P_{up}$ and f) $P_{down}$, and 50 u.c. g) $P_{up}$ and h) $P_{down}$. The measured magnetic field is marked at the top left corner.

Typical results of reversible control (two $P_{up}$/$P_{down}$ circles) of transport properties for BTO (10 u.c.)/LSMO (15 u.c.) are shown in Figure S7, where the angular MR measurements were carried out under a magnetic field of 5 T at 10 K. A series of gate voltage of –4 V → +4 V → –4 V → +4 V are applied on the sample to alter its polarization state as follows: $P_{up}$-1 → $P_{down}$-1 → $P_{up}$-2 → $P_{down}$-2. The two measurements of the same polarization state show good consistency, both in the temperature dependence of the resistance (Figure S7a) and angular MR curves (Figure S7b and S7c), suggesting that the electrical manipulation of transport properties is reversible and reliable.

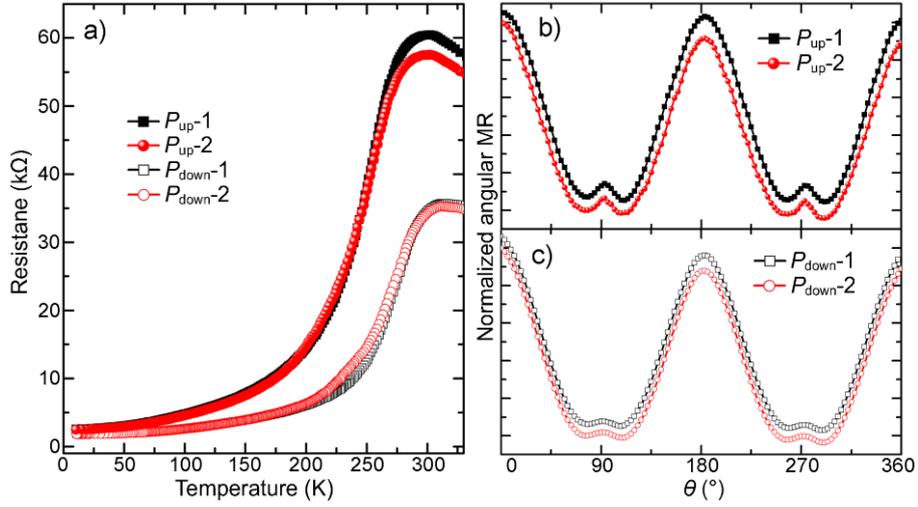

**Figure S7** Reversible control of a) temperature dependence of the resistance and angular MR for BTO (10 u.c.)/LSMO (15 u.c.) under b) $P_{up}$ and c) $P_{down}$ state. The angular MR measurements were carried out under a magnetic field of 5 T at 10 K. A series of gate voltage of $-4$ V $\rightarrow$ $+4$ V $\rightarrow$ $-4$ V $\rightarrow$ $+4$ V are applied on the sample to alter its polarization state as follows: $P_{up}$-1 $\rightarrow$ $P_{down}$-1 $\rightarrow$ $P_{up}$-2 $\rightarrow$ $P_{down}$-2.

**Original XAS and XLD signals for all the samples**

The XAS and corresponding XLD are currently of great interest due to its promise of providing the ground-state electronic structure and orbital ordering with a high element resolution. The XAS spectra of Mn were normalized by dividing the spectra by a factor such that the $L_3$ pre-edge and $L_2$ post-edge have identical intensities for the two polarizations. After that, the pre-edge spectral region was set to zero and the peak at the $L_3$ edge was set to one. XLD, the difference between the two measurements ($I_{ab}$–$I_c$), gives direct insight of the relative occupancies of $3z^2 - r^2$ [$P(3z^2 - r^2)$] and $x^2 - y^2$ [$P(x^2 - y^2)$] orbital: the more negative (positive) $A_{XLD}$ is, the larger relative occupancy of $d_{x^2-y^2}$ ($d_{3z^2-r^2}$) is.[S6]

The comparison of XLD in Figure S8 presents the information for the orbital occupancy of BTO/LSMO with different LSMO thicknesses under $P_{up}$ and $P_{down}$. The $A_{XLD}$ is negative for the case of $P_{up}$, suggesting that the preferential orbital occupancy of LSMO is $x^2 - y^2$, while BTO of $P_{down}$ favors the $3z^2 - r^2$ occupancy, regardless of the LSMO thickness. The variation in orbital occupancy is direct clue for the modulation of orbital hybridization under FE polarization. Note that the proportion of $3z^2 - r^2$ orbital increases with the LSMO thickness as the tensile strain from substrate is gradually relaxed.

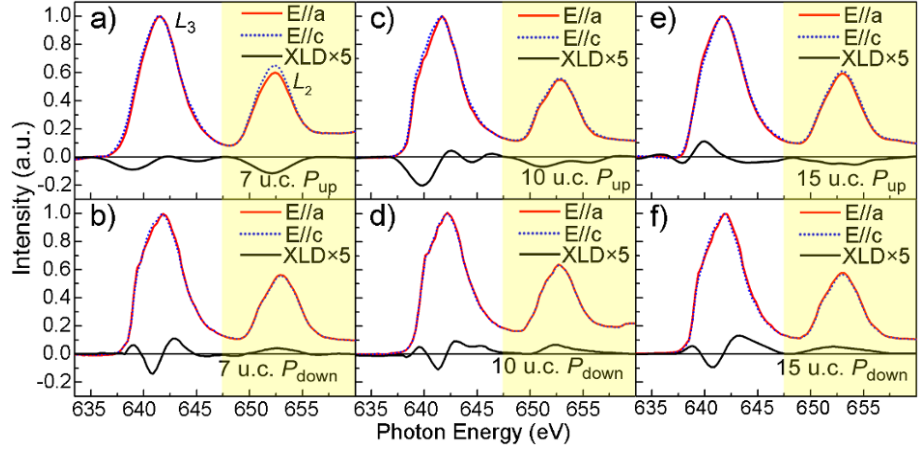

**Figure S8** Normalized XAS and XLD (the intensity is multiplied by 5 times) of BTO/LSMO heterostructures with different LSMO thicknesses and polarization states: 7 u.c. a) $P_{up}$ and b) $P_{down}$, 10 u.c. c) $P_{up}$ and d) $P_{down}$, and 15 u.c. e) $P_{up}$ and f) $P_{down}$.

### The control experiment carried out on the (110) heterostructure

The preparation of BTO/LSMO heterostructures on (110) STO in this control experiment is identical with the (001) BTO (10 u.c.)/LSMO (15 u.c.), as they were grown simultaneously in the same chamber. The (110) heterostructure exhibits obvious FE properties according to the out-of-plane PFM image in Figure S9a, although the contrast of this image is not as sharp as that of (001) one. However the sketch of (110) heterostructure in Figure S9b suggests that the FE polarization could not affect the orbital hybridization between Ti and Mn due to the absence of a key bridge of O ion. Thus the FE polarization could not influence the orbital-related magnetic and electric properties of the heterostructure.

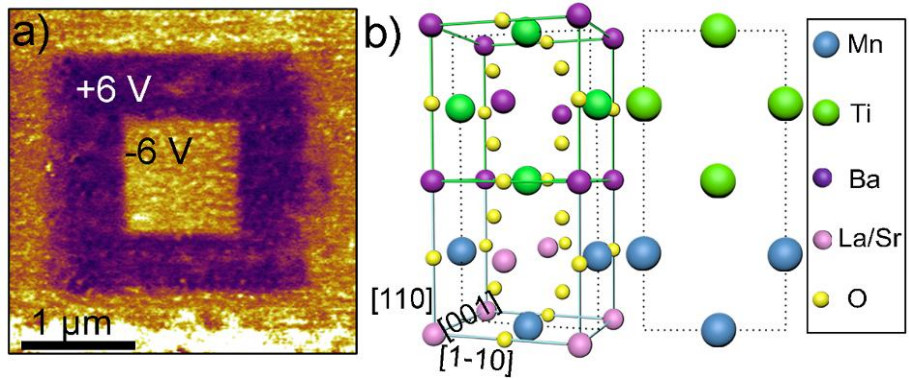

**Figure S9** a) Typical out-of-plane PFM image of BTO/LSMO on (110) STO. b) Sketch for atomic arrangement of heterostructure on (110) STO and the cross-section view along (002) plane.

In contrast to the case of (001) sample, the resistance and Curie temperature ($T_C$) of (110) heterostructures are almost independent of the polarization state of BTO as shown in Figure S10a. A gate voltage of ±6 V was used to switch the polarization in (110) heterostructures. It is mainly because that the stoichiometric of LSMO is $La_{2/3}Sr_{1/3}MnO_3$, at the center of FM phase region in the phase diagram, which should not be sensitive to the merely small changes of carrier density without variation of orbital occupancy under different polarization states. The (110) heterostructures exhibit stronger pMR peaks at 90° and 270° compared with the case of (001) in Figure S10b and S10c, as the interfacial in-plane $x^2 - y^2$ orbital occupancy is enhanced without out-of-plane orbital hybridization. Furthermore, the pMR peaks are stable as the polarization state of BTO switches from $P_{up}$ to $P_{down}$. These results obtained from (110) heterostructure illustrate that the importance of orbital modulation in controlling the magnetism of metallic LSMO by FE polarization. Although the lattice distortions in (001) and (110) heterostructures might be somehow different, the magnetic and conductive properties of these two systems at $P_{down}$ states are quite close (Figure S10a). The carrier density modulated by vertical upward/downward polarization (Figure S9a), if any, should be the same for the two samples. Thus the control experiment provides additional evidence that orbital reconstruction accounts for the ME coupling.

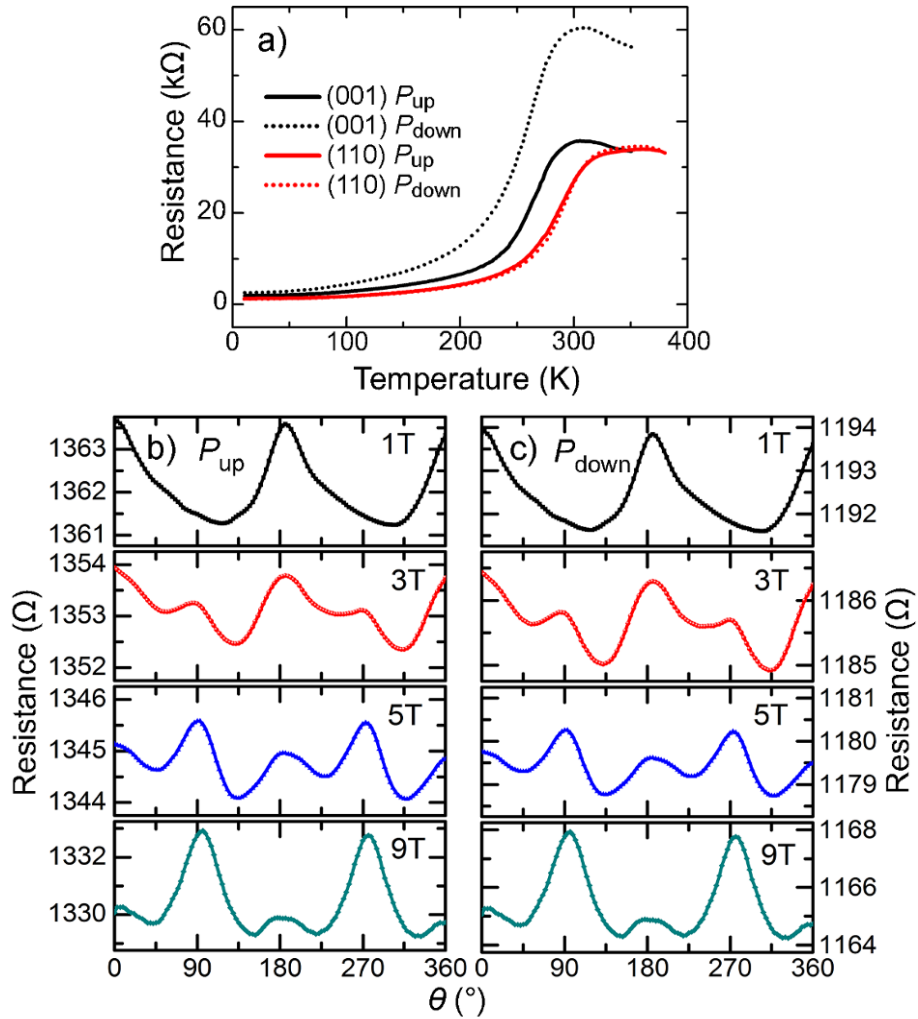

**Figure S10** a) The temperature dependence of the resistance curves of (001) and (110) heterostructures with different polarization states. The angular MR for (110) BTO/LSMO under b) $P_{up}$ and c) $P_{down}$ states.